\documentclass[preprint,12pt]{elsarticle}

\usepackage{amssymb}\usepackage{multirow}
\usepackage{amsmath}
\usepackage{xcolor}

\journal{Physica B: Condensed Matter}

\begin{document}

\begin{frontmatter}



\title{Self-oscillations induced by self-induced torque in magnetic double tunnel junction} 

\author[srmtrp,eas]{R.~Arun \corref{cor1}}
\ead{arunbdu@gmail.com}
\author[sas]{R.~Gopal}
\ead{gopalphysics@gmail.com}
\author[sas]{V.~K.~Chandrasekar}
\ead{chandru25nld@gmail.com}
\author[cnld]{M.~Lakshmanan}
\ead{lakshman.cnld@gmail.com}

\cortext[cor1]{Corresponding author}

\address[srmtrp]{Center for Nonlinear and Complex Networks, SRM TRP Engineering College, Tiruchirappalli - 621 105, India}
\address[eas]{Center for Research, Easwari Engineering College, Chennai - 600 089, India}
\address [sas]{Centre for Nonlinear Science \& Engineering, School of Electrical \& Electronics Engineering, SASTRA University, Thanjavur - 613 401, India.} 
\address[cnld]{Department of Nonlinear Dynamics, School of Physics, Bharathidasan University, Tiruchirapalli - 620 024, India}

%
%
\begin{abstract}
Self-oscillations of the magnetization due to self-induced torque (SIT) in a magnetic double tunnel junction that consists of perpendicularly polarized, pinned and free layers is investigated along with the field-like torque (FLT). The associated Landau-Lifshitz-Gilbert-Slonczewski equation is numerically analysed to exhibit the oscillations of magnetization driven by the current. From the numerical analysis, we show that the SIT is essential to generate oscillations in the order of GHz and without it the magnetization reaches steady state after exhibiting switching. Without FLT, the frequency of the oscillations decreases with the current while the power of oscillations increases. In the presence of the negative strength of the FLT the power spectral density confirms that the frequency, power and the Q-factor increase with the current. Also the tunability range and the rate at which the frequency enhances increase with the magnitude of the FLT.
\end{abstract}




\end{frontmatter}



\section{Introduction}
The oscillations of magnetization in spin-torque nano oscillators is a fascinating phenomenon and has applications for the generation of microwave oscillations \cite{kiselev,locatelli,chen,pmgunn,jiang,kubler}. The oscillations of the magnetization are converted into voltage oscillations via a magnetoresistance effect, such as Giant Magnetoresistance (GMR) or Tunneling Magnetoresistance (TMR)~\cite{zeng}. The external magnetic field causes the magnetization vector to precess around the direction of the field and the damping within the system makes it dissipate the energy, consequently the magnetization settles along the field axis.  The dissipation can be compensated by the energy due to the spin-transfer torque (STT)~\cite{ralph} exerted by the current applied to the system and the magnetization can be made to precess continuously when the energy supplied by spin-transfer torque compensates for the intrinsic damping losses \cite{kim}. The frequency of the magnetization oscillations can be tuned in the order of GHz by the field and current~\cite{arunprb,arunjmmm}. 

The sustained periodic oscillations driven solely by a constant current (dc) without any periodic inputs, which is referred as self oscillation \cite{taniguchi_apl, taniguchi_jap} is essential in applications which avoid magnetic fields or periodic driving forces. It is possible to tightly integrate arrays of STNO neurons on top of CMOS circuits without the need for an external magnetic field. This greatly streamlines the design process and encourages brain-like parallelism in hardware neuromorphic computing systems \cite{bohnert}. A stable self-oscillations can be excited in  spin-torque nano oscillator by the presence of field-like torque (FLT). Its presence modifies the energy balance between the STT and the damping \cite{taniguchi_apl,taniguchi_jap,guo}.

Recently, a new type of torque denoted as self-induced torque has been proposed in a magnetic double tunnel junction (MDTJ) due to the spins pumped by the magnetization rotations in the free layer \cite{gunnink}. The current and/or field rotate the magnetization and this rotation pumps the spins into the two adjacent normal metal layers within the system \cite{berkov1}. Further these spins alter the current flowing through the system. The STT induced by these pumped spins is called self-induced torque (SIT). Due to the discrepancy between the theory and experiments the self-induced torque is essentially incorporated in the Landau-Lifshitz-Gilbert-Slonczewski (LLGS) equation \cite{xiao,tiber,berkov2}.

In the case of current driven STNO, Taniguchi $et~al$ \cite{taniguchi_jap} have reported that self-oscillations are possible due to FLT and Gunnink $et~al$ \cite{gunnink} have formulated SIT in MDTJ. However, the impact of this newly formulated SIT on the self-oscillations of MDTJ is not yet understood. In this paper, we demonstrate that self-oscillations in MDTJ can be induced by SIT and frequency of the oscillations can be tuned and enhanced by FLT.
We numerically study the temporal evolution of the magnetization in MDTJ driven by  current in the presence of FLT  and SIT. We explore the oscillations of the magnetization with frequency in the range of  GHz. We demonstrate that self-oscillation of the magnetization is possible by the mere presence of SIT within this macrospin model and in the absence of the SIT there exists only switching. Without FLT the tunability of the frequency by current is considerably small and in the presence of FLT we show an increment in the tunability of frequency as well as an enhancement in the range (or amplitude) of the oscillations. The frequency linearly increases and decreases with current for the negative and positive FLT strength, respectively. For the negative FLT strength the frequency, power and the Q-factor increase with current.

The paper is organized as follows: Sec.2 covers the model and the governing equation and in Sec.3 the emergence of the self-oscillations due to the SIT is discussed. The impact of the FLT on the frequency and power of the self-oscillations is investigated in Sec.4. Finally the results are summarized in Sec.5. In Appendix A the component-wise  derivation of the magnetization is presented.

\section{Model}
\begin{figure}[h!]
	\centering	\includegraphics[width=0.4\linewidth]{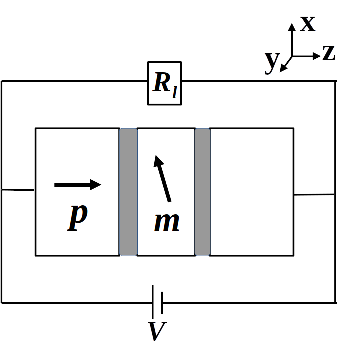}
	\caption{ Schematic diagram of a system of magnetic double tunnel junction. The free layer in the middle is tunnel coupled with a pinned layer at left and normal metal layer at right.}
	\label{Fig1}
\end{figure}

The schematic picture of the MDTJ which is displayed in Fig.\ref{Fig1} consists essentially of three layers: (i) pinned layer at the far left, (ii) normal metal layer at the far right and (iii) both these are tunnel coupled with the free layer at the middle. The free and the pinned layers are ferromagnetic layers. In the free layer the direction of the magnetization can change, while this cannot happen in the pinned layer. In the pinned layer the magnetization is fixed in a particular direction. The unit vectors of the magnetization in the free and pinned layers are denoted by ${\bf m}$ and ${\bf p}$, respectively. The coordinate axes are shown in Fig.\ref{Fig1}, where the unit vectors along the positive $x$, $y$, and $z$ axes are denoted by ${\bf e}_x$, ${\bf e}_y$, and ${\bf e}_z$, respectively.  The magnetization of the pinned layer is taken as ${\bf p}$ = ${\bf e}_z$.

The LLGS equation, along with the self-induced torque obtained through the charge conservation that governs the dynamics of the magnetization in the free layer, is given by \cite{gunnink}
\begin{align}
	\frac{d\bf m}{dt} = &-\gamma ~ {\bf m}\times{\bf H}_{eff} + \alpha ~{\bf m}\times {\frac{d\bf m}{dt}}+\gamma H_s~ {\bf m}\times({\bf m \times \bf p}) \nonumber\\&+\gamma \beta H_s {\bf m}\times {\bf p}	+\gamma ~ G R ~{\bf m\times (m\times p)}, \label{llgs}
\end{align}

where $R = {\bf p}\cdot ({\bf m \times \frac{d\bf m}{dt}})$.  The first and second terms at the right hand side of the Eq.\eqref{llgs} are responsible for the precession and damping of the magnetization ${\bf m}$, respectively.  The third, fourth, and fifth terms are the STT, FLT, and SIT, respectively.  In Eq.\eqref{llgs}, the quantity ${\bf H}_{eff}$ is the effective field, $\gamma$ is the gyromagnetic ratio, $\alpha$ is the damping parameter, $H_s$ the strength of the STT given by 
\begin{eqnarray}
	H_s = \frac{\hbar \eta I}{2eM_s  W(1+\lambda {\bf m}\cdot {\bf p})}. \label{Hs}
\end{eqnarray}
Here, $\hbar(=h/2\pi)$ is the reduced Planck's constant, $e$ is the electron charge, $M_s$ is the saturation magnetization, $\eta$ and $\lambda$ are the dimensionless parameters which determine the magnitude and angular dependence of the spin-transfer torque, $I$ is the current passing through the system, and $W$ is the volume of the free layer.  The effective field ${\bf H}_{eff}$ consists of the anisotropy field and the demagnetization field and it is given as 
\begin{eqnarray}
{\bf H}_{eff} = H_k m_z {\bf e}_z- 4\pi M_s m_z {\bf e}_z. \label{heff}
\end{eqnarray}
Here $H_k$ is the anisotropy field and $M_s$ is the saturation magnetization. 

The damping parameter $\alpha$ is given by 
\begin{eqnarray}
\alpha = \alpha_0+ \frac{\hbar^2\gamma}{4e^2 M_s W} [\tilde{g}_l+\tilde{g}_r], \nonumber
\end{eqnarray}
where  $\tilde{g}_l$ and $\tilde{g}_r$ are the spin-flip conductances of the left and right junctions, respectively, $\alpha_0$ accounts for the internal Gilbert damping \cite{gilbert}.

In Eq.\eqref{llgs}, $\beta$ is the strength of the field-like torque, and $G$ is the strength of the SIT.
Here, $G=\mu H_s$, where $\mu$ is the ratio between the SIT and the STT.  $\mu$ is defined as $\mu = -{\hbar g_l^s}/{4 e g_r V}$, where $V$ is the bias voltage applied across the whole system\cite{gunnink}. $g_l^s$ is the spin conductance of the left junction and $g_r$ is the charge conductance of the right junction. Voltage $V$ is given by 
\begin{eqnarray}
	V = \frac{I {R_l (R_0-\triangle R~ m_z)}}{({R_l+R_0-\triangle R~ m_z})}. \nonumber
\end{eqnarray}
$R_l$ is the load resistance, $R_0 = R_p+\triangle R $, and $\triangle R$ = 0.5 $R_p G_{MR}$. $R_p$ is the resistance when the magnetization of the free layer is parallel to that of the pinned layer, and $G_{MR}$ is the giant magneto-resistance. Here,  the   terms $\tilde{g}_l$, $\tilde{g}_r$, $g_l^s$ and $g_r$ are defined as\cite{gunnink}
\begin{eqnarray}
	\tilde{g}_l &=& \frac{1}{2}[g_l^P+g_l^{AP}-(g_l^P-g_l^{AP}){\bf m\cdot p}], \nonumber\\
	\tilde{g}_r &=& g_r = g_l^P,~~~g_l^s = P^{-1}(g_l^P-g_l^{AP}). \nonumber
\end{eqnarray}
In the above, $g_l^P$ and $g_l^{AP}$ are the conductances of the left side junction when the magnetization of the free layer is parallel and anti-parallel to the fixed layer's magnetization, respectively. $P$ is the polarizing factor of the nanomagnet. After incorporating $H_{s}$ and $\mu$, $G$ can be obtained as
\begin{align}
	G=\frac{\hbar^2 \eta ~({R_l+R_0-\triangle R~ m_z})}{8 e^{2} M_{e} W (1+\lambda ~m_{z})  {R_l (R_0-\triangle R~ m_z)}} \frac {g_l^s}{g_r}. \nonumber
\end{align}

\begin{figure}[h!]
	\centering	\includegraphics[width=0.8\linewidth]{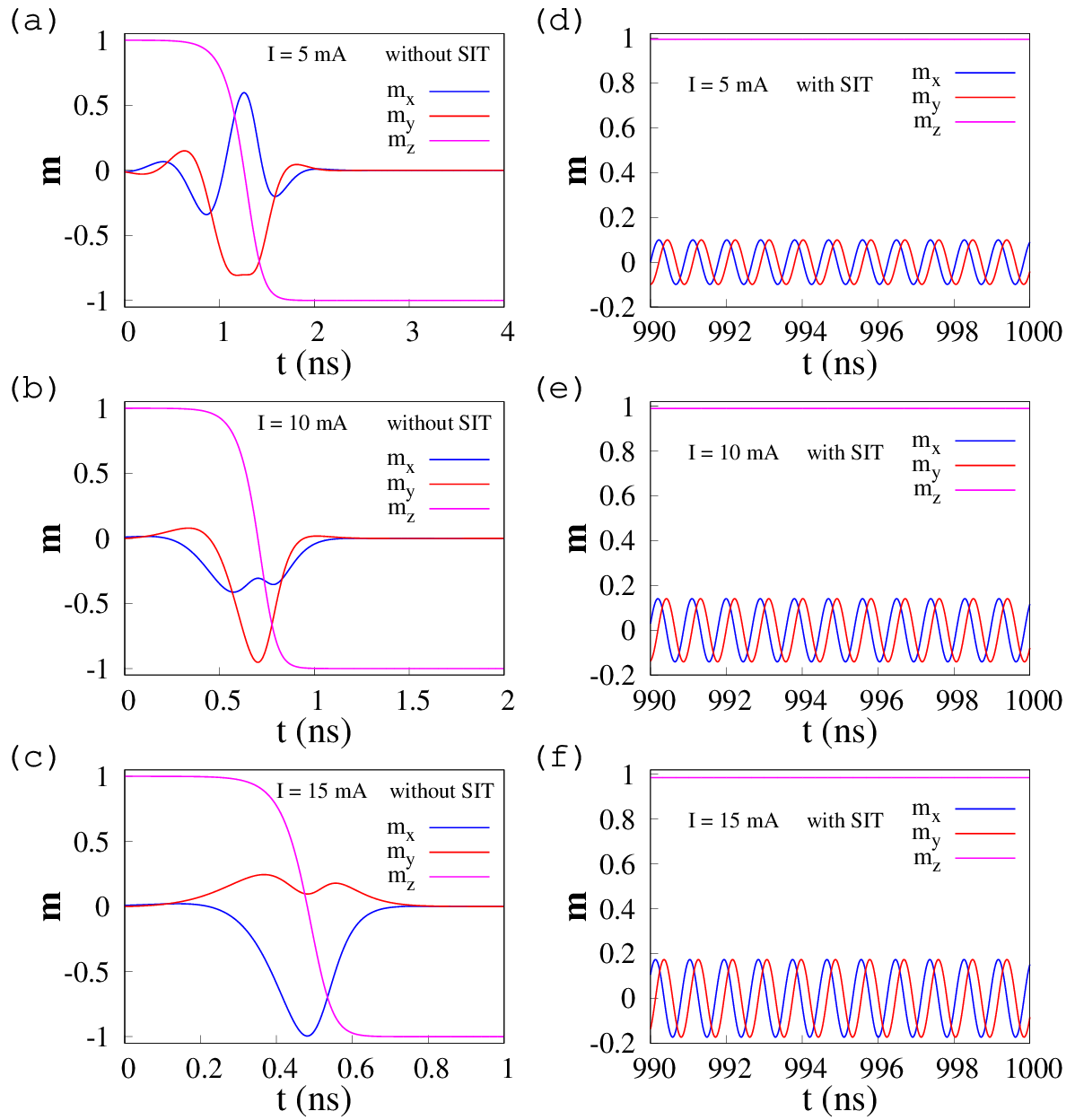}
	\caption{Oscillation of the magnetization components $m_x$, $m_y$, and $m_z$ in the (a-b) absence and (d-f) presence of the SIT corresponding to the currents $I$ = 5, 10, and 15 mA.}
	\label{Fig2}
\end{figure}

The temporal variation in the direction of the magnetization of the free layer is analysed through macrospin approximation \cite{ripp,kubo,Tani} by numerically solving Eq.\eqref{llgs} using by  adaptive-step-size Runge-Kutta fourth order method after an adequate sampling time with the parameters  \cite{taniguchi_apl,gunnink}
listed in Table.\ref{table1}. All the computations given below have been carried out from the time series data computed upto 1000 ns with the time step not greater than $10^{-12}$ s.
\begin{table}[htbp!]
\begin{center}
\caption{\label{tab:table1} Values of different parameters, selected Refs.\cite{taniguchi_apl,taniguchi_jap,gunnink,slon}}.
\begin{tabular}{|c|l|c|l|} 
\hline
{\bf Parameters} &  {\bf Values} & {\bf  Parameters }&  {\bf Values}\\ 
\hline
$M_s$ &  1.4483$\times 10^6$ Am$^{-1}$& $h$ &  6.62607$\times 10^{-34}$ J$\cdot$s \\ 
$H_k$ &  1.86 T &$e$ &  1.602$\times 10^{-19}$ C \\ 
$\gamma$ &  1.764$\times 10^{11}$ rad (Ts)$^{-1}$&$g_l^P$ &  0.12 $G_0$ \\ 
$\eta$ & 0.54&$g_l^{AP}$ &  0.05 $G_0$ \\
$\lambda$ & $\eta^2$&$G_0$ & 2$e^2$/h\\
$P$ & 0.4&$R_p$ & 100 $\Omega$\\
$W$ & $\pi \times 60^2 \times 2$ nm$^3$&$G_{MR}$ & 0.7 \\
$R_l$ & 50 $\Omega$& &\\
\hline
\end{tabular}
\label{table1}
\end{center}
\end{table}	

The initial conditions for ${\bf m}$ for computation are randomly taken near $m_x$ = 1 unless otherwise specified. Henceforth, the results corresponding to the absence of SIT will be obtained by setting $G$ = 0.

\section{Emergence of self-oscillations due to SIT}
\begin{figure}[h!]
	\centering	\includegraphics[width=0.6\linewidth]{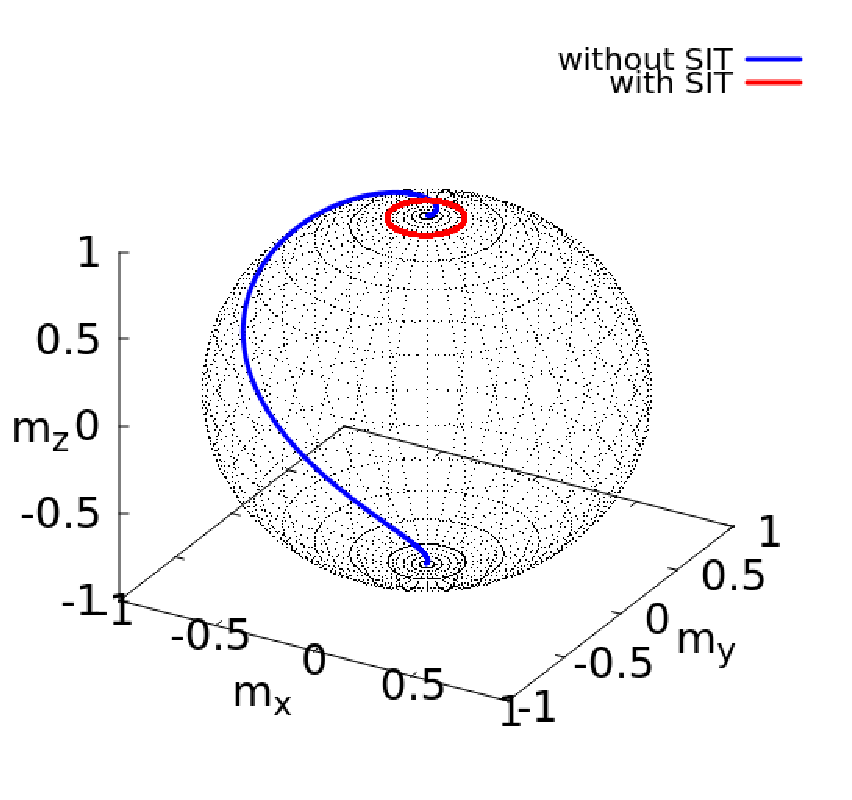}
	\caption{ Trajectory of the magnetization without (blue) and with (red) SIT for $I$ = 15 mA and $\beta$ = 0.}
	\label{Fig10}
\end{figure}
Initially, the nonexistence of the magnetization oscillations in the absence of SIT ($G=0$) and its emergence in the  presence of the SIT  ($G\ne 0$)  without the field-like torque ($\beta$ = 0) are shown in Figs.\ref{Fig2}(a-f).  Figs.\ref{Fig2}(a-c) show the steady state motion of the magnetization without any oscillations from $m_z$ = 1 to -1 for different values of currents $I$ = 5, 10, and 15 mA in the absence of SIT and Figs.\ref{Fig2}(d-f) confirm the oscillations in the presence of SIT. From Figs.\ref{Fig2}(a-c) we can understand that in the absence of SIT the magnetization switches its direction from $m_z$ = 1 to $m_z$ = -1. Without SIT oscillations are not possible even with a large magnitude of the current. Figs.\ref{Fig2}(d-f) indicate that oscillations are induced by the presence of SIT and also the amplitude of the oscillations is increased with the increase of the current. However we can also observe that the frequency of the oscillations slightly differs around 1 GHz for all the currents. The temporal evolution of $m_z$ in Figs.\ref{Fig2}(d-f) show that $m_z$ is nearly constant, which implies that the magnetization precesses in the $xy$-plane perpendicular to the $z$-axis. Figs.\ref{Fig2}(a-f) confirm that the emergence of the oscillations without any external magnetic field, i.e. zero-field oscillation, can be exhibited by incorporating the SIT.
\begin{figure}[h!]
	\centering	\includegraphics[width=1\linewidth]{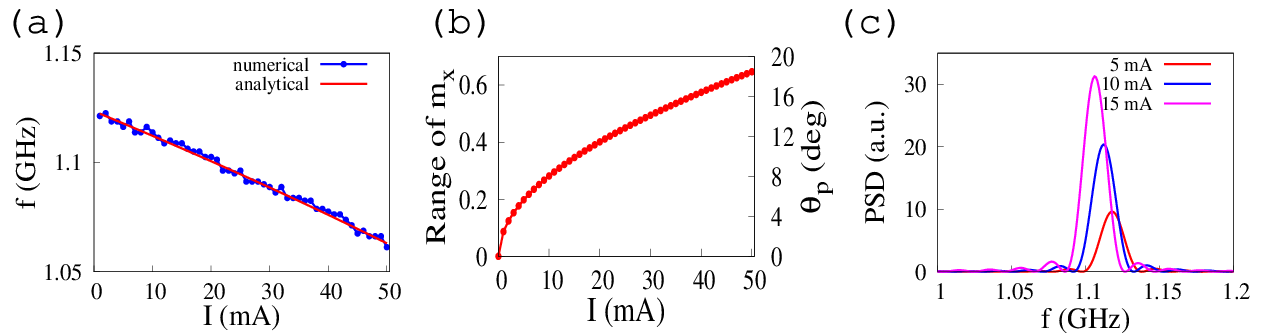}
	\caption{(a) Frequency and (b) range and precession angle $\theta_p$ of oscillations with respect to the current. (c) PSD for the currents $I$ = 5, 10, and 15 mA. Here, $\beta$ = 0.}
	\label{Fig3}
\end{figure}

The terms corresponding to STT and SIT are given by
\begin{align}
	\mathcal{T}_{STT} &= \gamma H_s {\bf m \times (m \times p)},\nonumber\\
	\mathcal{T}_{SIT} &= \gamma ~G~R~ {\bf m \times (m \times p)} \nonumber\\&= -\gamma H_s ~ \frac{ \hbar g_l^s}{4eg_r V} [{\bf p\cdot (m\times \dot m)}] ~ {\bf m \times (m \times p)}.\nonumber
\end{align}	
The trajectories of the magnetization without and with SIT for $I = 15$ mA and $\beta$ = 0 are plotted in Fig.\ref{Fig10} by blue and red solid lines, respectively. When the current is applied, the magnetization starts to precess around the positive $z$-axis. Since the direction of STT is away from the positive $z$-axis, the magnetization is switched to the opposite direction and settles along the negative $z$-axis in the absence of SIT as shown in Fig.\ref{Fig10}.  When the SIT is present, the direction of the torque exerted by it is exactly opposite to that of STT and hence the magnetization is not allowed to switch toward negative $z$-direction and both torques balance in such a way that the magnetization continuously precesses around the positive $z$-direction as observed in Fig.\ref{Fig10}. Therefore, without SIT the magnetization does not exhibit continuous precession around positive $z$-axis and is forced to switch along the negative $z$-axis due to STT. The SIT exerts torque to the opposite direction of STT and makes the magnetization to exhibit stable precession.

The frequency, range and power spectral density (PSD) in the presence of SIT are plotted in Figs.\ref{Fig3}(a-c), respectively. The frequency of oscillations against current is plotted in Fig.\ref{Fig3}(a) by blue line-points, where we can see that the range of tunability in the frequency is small and also the frequency reduces with an increase of the current. The non-smoothness in the frequency curve is due to the randomness in selection of the initial condition around $m_z$ = 1. The red solid line represents the frequency fitted from the relation which is derived as follows:
Since the precession of the magnetization takes place in the $xy$-plane as shown in Figs.\ref{Fig2}(d-f) we can substitute $\frac{dm_z}{dt}=0$ and $m_z = m_z^*$=constant corresponding to $t = \infty$ in Eq.\eqref{eq:sup8} in the Appendix A and we can obtain
\begin{align}
	\alpha (H_k-4\pi M_s) m_z^* - H_s(1+\alpha\beta +\mu B) = 0. \label{eq1}
\end{align}
Equations \eqref{eq:sup6} and \eqref{eq:sup7} can be written in terms of cartesian coordinate system with $m_x = \sin\theta\cos\phi, ~m_y = \sin\theta\sin\phi, ~m_z = \cos\theta$ as
\begin{align}
	(1+\alpha^2)&[\cos\theta\cos\phi\frac{d\theta}{dt}-\sin\theta\sin\phi\frac{d\phi}{dt}] = -(H_k-4\pi M_s)\cos\theta(\sin\theta\sin\phi+\alpha\cos\theta\sin\theta\cos\phi) \nonumber\\
	&+H_s \cos\theta\sin\theta\cos\phi(1+\alpha\beta+\mu B)+H_s \sin\theta\sin\phi(\beta-\alpha-\alpha\mu B), \label{eq2}\\
	(1+\alpha^2)&[\cos\theta\sin\phi\frac{d\theta}{dt}-\sin\theta\cos\phi\frac{d\phi}{dt}] = -(H_k-4\pi M_s)\cos\theta(\sin\theta\cos\phi-\alpha\cos\theta\sin\theta\sin\phi) \nonumber\\
	&+H_s \cos\theta\sin\theta\sin\phi(1+\alpha\beta+\mu B)-H_s \sin\theta\cos\phi(\beta-\alpha-\alpha\mu B). \label{eq3}
\end{align}
Adding Eqs.\eqref{eq2} and \eqref{eq3} after multiplying them with $\sin\phi$ and $\cos\phi$, respectively, we can get
\begin{align}
	\frac{1+\alpha^2}{\gamma} \frac{d\phi}{dt} = (H_k-4\pi M_s) \cos\theta - H_s(\beta-\alpha-\alpha\mu B).\label{eq4}
\end{align}
At precession $\frac{d\phi}{dt}~=~2\pi f$, where $f$ is the frequency, and $\cos\theta = \cos\theta^*=m_z^*$=constant and by substituting these along with Eq.\eqref{eq1} in Eq.\eqref{eq4} we can obtain the relation for fitting the frequency
\begin{align}
	f = \frac{\gamma}{2\pi \alpha}H_s (1+\mu B)\label{freq},
\end{align}
where 
\begin{align}
	H_s = \frac{\hbar\eta I}{2eM_s W(1+\lambda m_z^*)},~~~~~~~~~ B = \frac{\gamma H_s (1-m_z^{*2})}{\alpha-\gamma H_s (1-m_z^{*2})}.
\end{align}

The frequency of the precession can be obtained from Eq.\eqref{freq} after substituting the value of $m_z^*$ obtained from numerical results. The frequency is  fitted by open circles in Fig.\ref{Fig3}(a) against the current. From Fig.\ref{Fig3}(a) we can see that the frequencies obtained from the relation Eq.\eqref{freq} (open circles) fit  well with the numerically plotted frequencies (solid lines). This implies the reliability of the numerical analysis in this work.

Figure \ref{Fig3}(b) confirms the increase in the range (difference between maximum and minimum values of the oscillation of $m_x$)  as well as the angle of precession $\theta_p$ around $z$-axis. At larger currents, the range increases linearly with the current. The PSD is plotted in Fig.\ref{Fig3}(c) for the three currents $I$ = 5, 10, and 15 mA, where we can see that the power increases with the magnitude of the current. The Q-factor, the ratio between the peak frequency and line-width, is calculated as 62.39, 61.47, and 61.21 for the currents 5, 10, and 15 mA, respectively. This conveys the increase in power while increasing the current without considerable change in the Q-factor. 
\begin{figure}[h!]
	\centering	\includegraphics[width=1\linewidth]{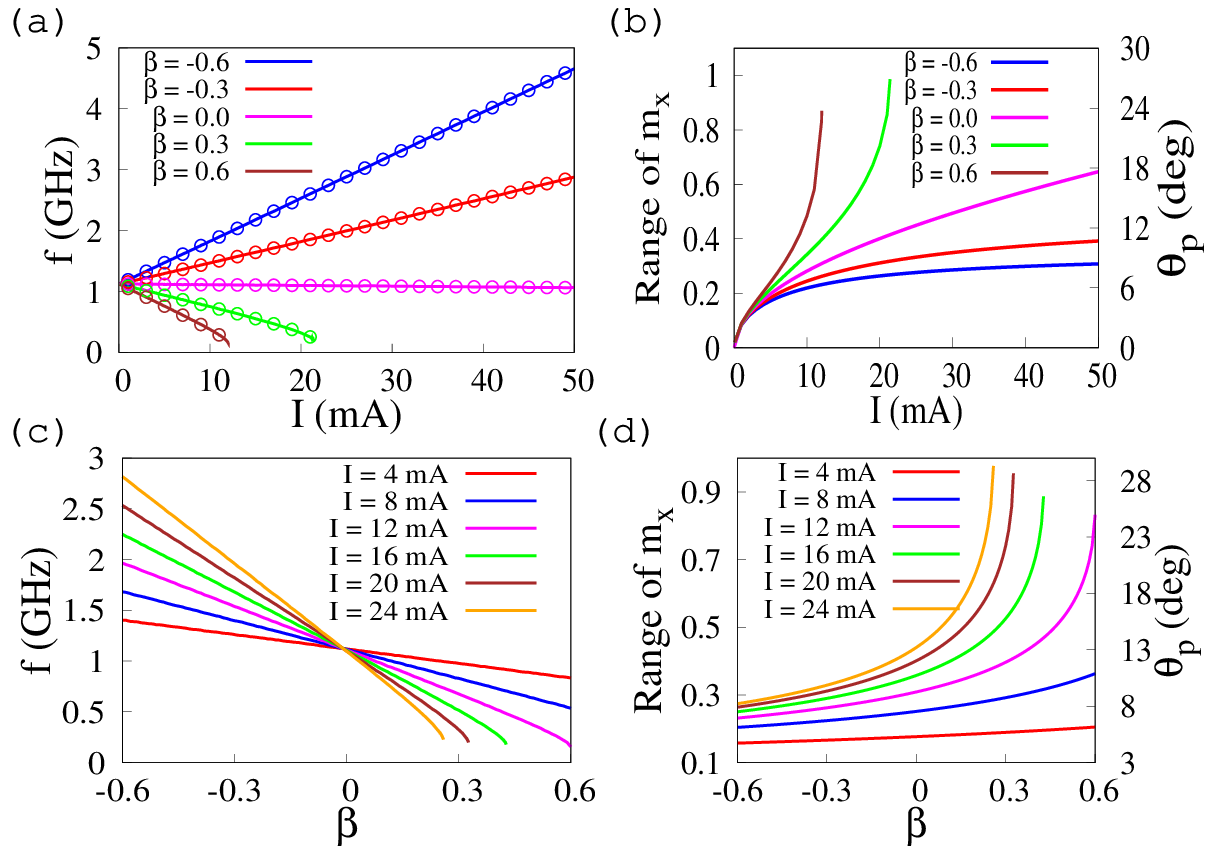}
	\caption{(a) Frequency and (b) range  and angle of precession $\theta_p$ of the oscillations for $\beta$ = -0.6 (blue), -0.3 (red), 0 (magenta), 0.3 (green), and 0.6 (brown)  against current. (c) Frequency and (d) range  and angle of precession $\theta_p$ of the oscillations for the currents $I$ = 4 mA (re), 8 mA (blue), 12 mA (magenta), 16 mA (green), 20 mA (brown), and 24 mA (orange) against the strength of FLT $\beta$. The open circles in (a) correspond to the frequency of oscillations obtained from Eq.\eqref{freq}.}
	\label{Fig4}
\end{figure}
\section{Impact of the FLT on the oscillations}
As we observe in Fig.\ref{Fig3} the increase in the current not only makes the considerable tunability in frequency but also decreases the  frequency.  Here we demonstrate that the frequency can be enhanced with the current in the presence of the FLT. In Fig.\ref{Fig4}(a) the frequency is plotted against the current for different strengths of the FLT, $\beta$ = -0.6, -0.3, 0, 0.3 and 0.6. The figure clearly demonstrates that the frequency increases and decreases linearly with current for negative and positive strengths of $\beta$, respectively. Also the rate of change of frequency with current increases with the magnitude of $\beta$.  For positive values of $\beta$ there exists a critical value for the current above which there is no oscillation as shown by the green and brown plots corresponding to the values of $\beta$ = 0.3 and 0.6, respectively. Above appropriate critical value of current, the magnetization exhibits only switching, from $m_z$ = 1 to -1, instead of oscillations. The critical value of current decreases with the increase of $\beta$. The corresponding range as well as angle of precession $\theta_p$ of the oscillations are shown in Fig.\ref{Fig4}(b) for different values of $\beta$.  This implies that the range  and angle of precession increase with the current upto the critical value for the positive strength of  FLTs. However, for the negative strength of FLTs the range of the oscillations increases and saturates at larger currents as shown in Fig.\ref{Fig4}(b). We can observe from Fig.\ref{Fig4}(a) and (b) that unlike the case $\beta$ = 0 (Fig.\ref{Fig3}(a)) where the range decreases while the frequency increases, here both the range and the frequency increase with current due to the presence of FLT ($\beta > $0).  In Fig.\ref{Fig4}(c) and (d) the frequency and the range of the oscillations are plotted against the strength of the FLT ($\beta$) for different values of the current $I$ = 4, 8, 12, 16, 20, and 24 mA. Figure \ref{Fig4}(c) implies that the frequency linearly increases and decreases with the magnitude of $\beta$ when $\beta <0$ and $\beta>0$, respectively. The rate of change of the frequency with FLT increases with the magnitude of the current. For $\beta>0$, there is a critical value for $\beta$ above which the magnetization exhibits no oscillations but switching. The critical value of $\beta$ decreases with the increase of the current. Figure \ref{Fig4}(d) evidences that the range  and angle of precession of oscillations increase with the strength of FLT for any current. However the rate at which the range increases is comparatively smaller for negative strength of the FLT when compared to positive strength of the FLT. 
\begin{figure}[h!]
	\centering	\includegraphics[width=0.5\linewidth]{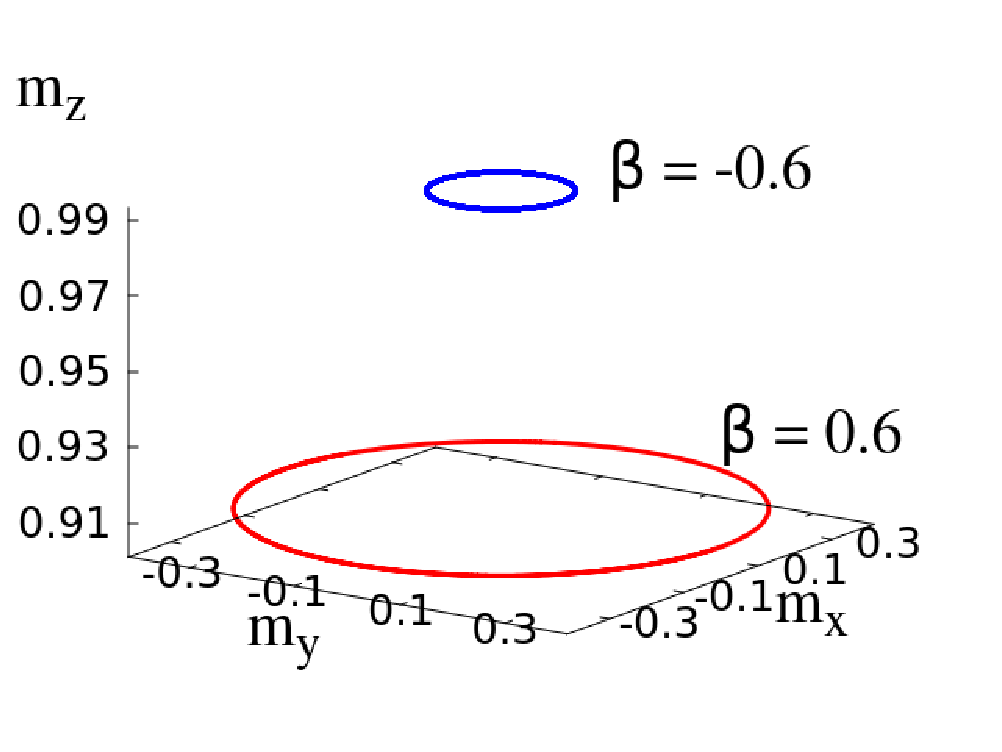}
	\caption{Trajectory of the magnetization in the $xy$-plane with precession for $\beta$ = 0.6 and -0.6. Here, $I$ = 12 mA.}
	\label{Fig5}
\end{figure}
Since the precession of the magnetization takes place in the $xy$-plane as shown in Fig.\ref{Fig5} the frequency of the precession is analytically determined from Eq.\eqref{freq}. The frequency is plotted by open circles in Fig.\ref{Fig4}(a) corresponding to different values of $\beta$. From Fig.\ref{Fig4}(a) we can see that the analytically obtained frequencies (open circles) match well with the numerically plotted frequencies (solid lines). 

\begin{figure}[h!]
	\centering	\includegraphics[width=0.5\linewidth]{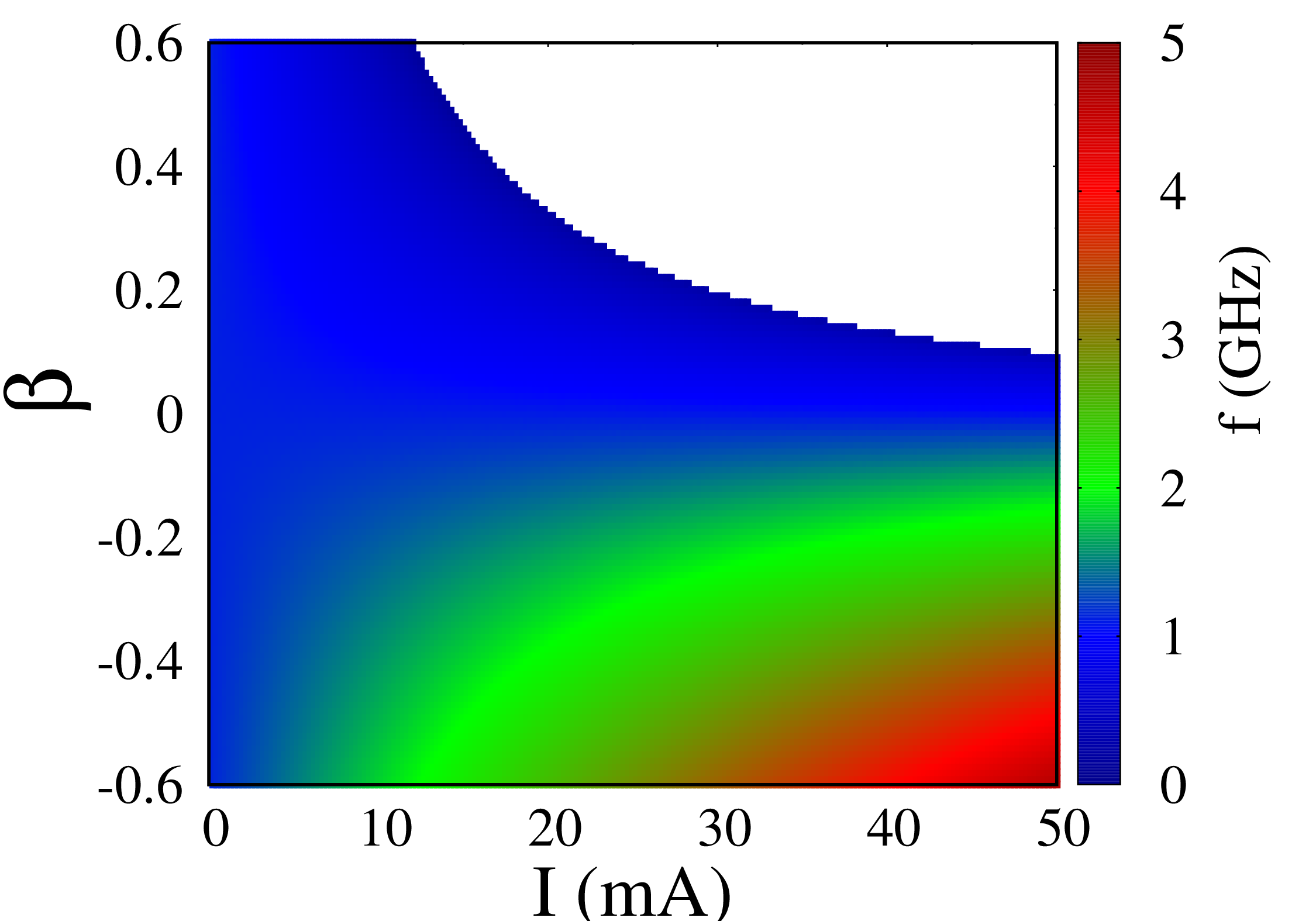}
	\caption{Frequency against current and $\beta$. White region is where the oscillations do not occur.}
	\label{Fig8}
\end{figure}

To verify the presence and absence of the oscillations and corresponding frequency for different pairs of values of current and $\beta$, we plot Fig.\ref{Fig8} for the frequency with respect to current and strength of the FLT $\beta$. From Fig.\ref{Fig8}, we observe that the oscillations are absent when the current and $\beta$ ($\beta >$0) are at their maximum values. Also, the frequency is maximum at the point where the current and $\beta$ ($\beta<0$) are maximum.

\begin{figure}[h!]
	\centering	\includegraphics[width=1\linewidth]{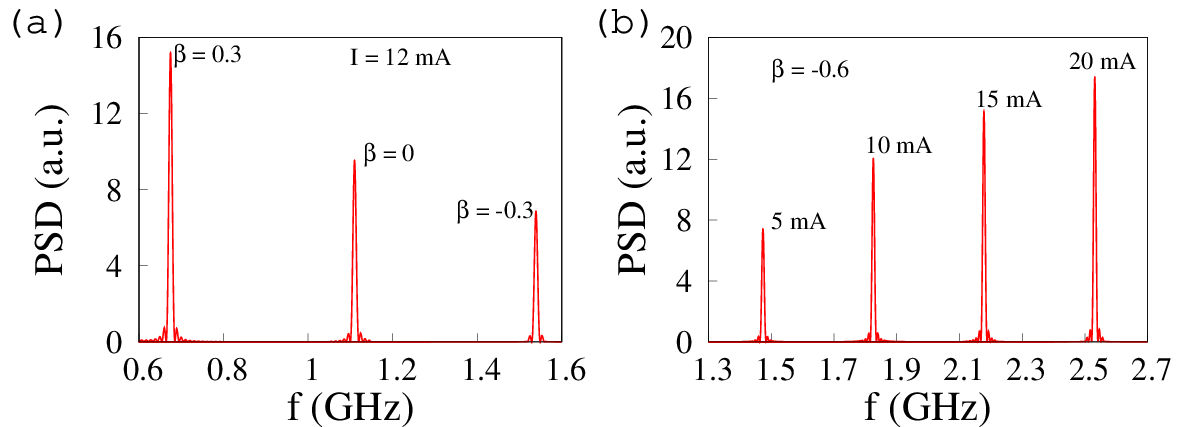}
	\caption{ PSD for different values of (a) the strength of FLT $\beta$ while $I$ = 12 mA and (b) current $I$ while $\beta$ = -0.6.}
	\label{Fig6}
\end{figure}

To verify the impact of FLT on PSD, Fig.\ref{Fig6}(a) and (b) are plotted for the PSD corresponding to  different values of $\beta$ as well as current from the time series data between $t$ = 900 ns and $t$ = 1000 ns. From Fig.\ref{Fig6}(a) it is clearly visible that the power decreases with the increase of frequency while varying the strength of FLT. The power for positive 
$\beta$ is comparatively larger than that of negative $\beta$.  
Figure \ref{Fig6}(b) confirms that the PSD increases with the current as well as the frequency for negative strength of FLT. 
The linewidth, frequency and Q-factor corresponding to the peaks given in Figs.\ref{Fig6}(a) and \ref{Fig6}(b) are listed in Table.\ref{table2} for the temperatures 0 K and 300 K. From Table.\ref{table2} we can understand that though the power decreases with the increase of frequency there is an enhancement in Q-factor with frequency. Also, there is an enhancement in frequency, power and Q-factor with current for negative FLT strength. Further, we can notice that the linewidth and Q-factor slightly decrease and increase, respectively, when the temperature is increased from 0 K to 300 K. The reason for this slight increment in Q-factor due to the thermal noise is attributed to the increase in the range of oscillations due to the temperature, as observed in Fig.\ref{Fig7}.

	\begin{table}
	\begin{center}
		\caption{\label{table2} Values of linewidth, frequency and Q-factor for different values of current and $\beta$.}
		\begin{tabular}{ |c|c|c|c|c|c|c|c| } 
			\hline
		\multirow{2}{*}{}	{\bf I (mA)} &  {\bf $\beta$} & \multicolumn{2}{|c|}{\bf  linewidth (MHz) }&  \multicolumn{2}{|c|}{\bf frequency (GHz)}&\multicolumn{2}{|c|}{\bf Q-factor}\\ 
			\cline{3-8}
			& & 0 K & {300 K}& 0 K & {300 K}& 0 K & {300 K}\\
			\hline
			12 &  -0.3 & 8.860 & {8.760}& 1.539& {1.540} &173.702 &{175.799}\\ 
			12 &  0.0 &8.860 & {8.770 }&  1.110& {1.116}&125.271 & {126.746}\\ 
			12 &  0.3 &9.058 & {8.685}&  0.675 & {0.678}&74.476  & {78.054}\\ 
			5 & -0.6&	8.880 &{8.730 }& 1.473&{1.476} &165.901 &{169.061}\\
			10 & -0.6 & 8.880& {8.790}&1.825& {1.828}&205.518 &{207.907}\\
			15 & -0.6 &8.860 &{8.810 }& 2.178&{2.181 }&245.813 &{247.560}\\
			20 & -0.6 &8.860 &{8.850 }& 2.531& {2.538}&285.711 &{286.249}\\
			\hline
		\end{tabular}
	\end{center}
\end{table}

We note that the present analysis is carried out within the macrospin approximation. While non-uniform magnetization modes may arise in perpendicular MTJ structures at sufficiently high current densities and large precession amplitudes, earlier experimental and theoretical studies have shown that, for the geometry and material parameters considered here, the macrospin description provides a reliable representation of the magnetization dynamics. In particular, experimental observations and macrospin-based modeling are reported in Refs.\cite{ripp,kubo,Tani} support the applicability of this approximation for similar systems. We therefore expect the main qualitative features of the self-oscillatory behavior discussed in this work to remain valid within the chosen framework, although a full micromagnetic treatment may be required to capture possible spatially non-uniform modes beyond the scope of the present study.

\section{In the presence of thermal noise}

Here, we investigate the impact of thermal noise on the oscillations. We include the thermal noise ${\bf H}_{th}$ in the effective field as follows and compute the frequency as we tune the current $I$ and $\beta$ in the presence of noise. 
\begin{align}
 {\bf H}_{eff} = H_k m_z {\bf e}_z - 4\pi M_s m_z {\bf e}_z + {\bf H}_{th},\nonumber
\end{align} 
where ${\bf H}_{th}$ is given by
\begin{align}
 {\bf H}_{th} = \sqrt{D}{\bf G},\hskip 1cm D = \frac{2\alpha k_b T}{\gamma M_s \mu_0 W \triangle t}. \nonumber
\end{align}
Here, ${\bf G}$ is the Gaussian random number generator vector of the oscillator with components $(G_x,G_y,G_z)$, which satisfies the statistical properties $<G_m(t)>=0$ and $<G_m(t)G_n(t')>=\delta_{mn}\delta(t-t')$ for all $m,n=x,y,z$. $k_B$ is the Boltzmann constant, $T$ is the temperature, $\mu_0$ is the magnetic permeability in free space, and $\triangle t$ is the step size of the time scale used in the simulation.

\begin{figure}[h!]
	\centering	\includegraphics[width=1\linewidth]{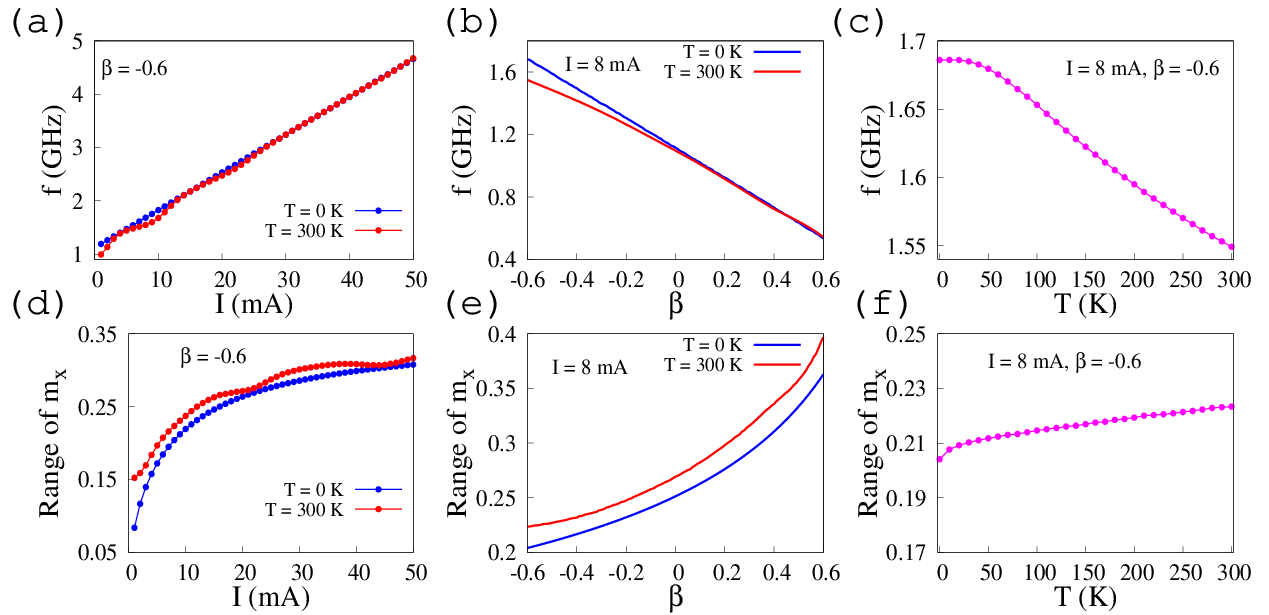}
	\caption{Frequency and range of oscillations computed for the temperatures $T$ = 0 and 300 K against (a,d) current $I$ for $\beta$ = -0.6 and (b,e) FLT strength $\beta$ for $I$ = 8 mA. (c,f) Frequency with respect to temperature $T$ for $\beta$ = -0.6 and $I$ = 8 mA.}
	\label{Fig7}
\end{figure}

We have plotted Fig.\ref{Fig7}(a-c) and (d-e) corresponding to the variations of the frequency and range of oscillations, respectively, against current $I$ for $\beta$ = -0.6, strength of the FLT $\beta$ for $I$ = 8 mA and the temperature for $\beta$ = -0.6 and $I$ = 8 mA, respectively. In Figs.\ref{Fig7}(a) and (b), the frequency is plotted for the temperatures $T$ = 0 K (blue) and 300 K (red), and we can observe that the frequencies are only slightly affected by the thermal noise. Fig.\ref{Fig7}(c) confirms that the frequency gradually reduces with the increase of temperature. The slight deviations in frequency due to thermal noise, as shown in Figs.\ref{Fig7}(a-c) imply that the system is stable against thermal noise.  From Figs.\ref{Fig7}(a) and (b), we can understand that the range of the oscillations is slightly more for the temperature $T$ = 300 K than 0 K and this means the range gets influenced by thermal noise. This is evidenced in Fig.\ref{Fig7}(c) also where we can see the slight increment in range due to the increase in temperature.

\begin{figure}[h!]
		\centering	\includegraphics[width=0.6\linewidth]{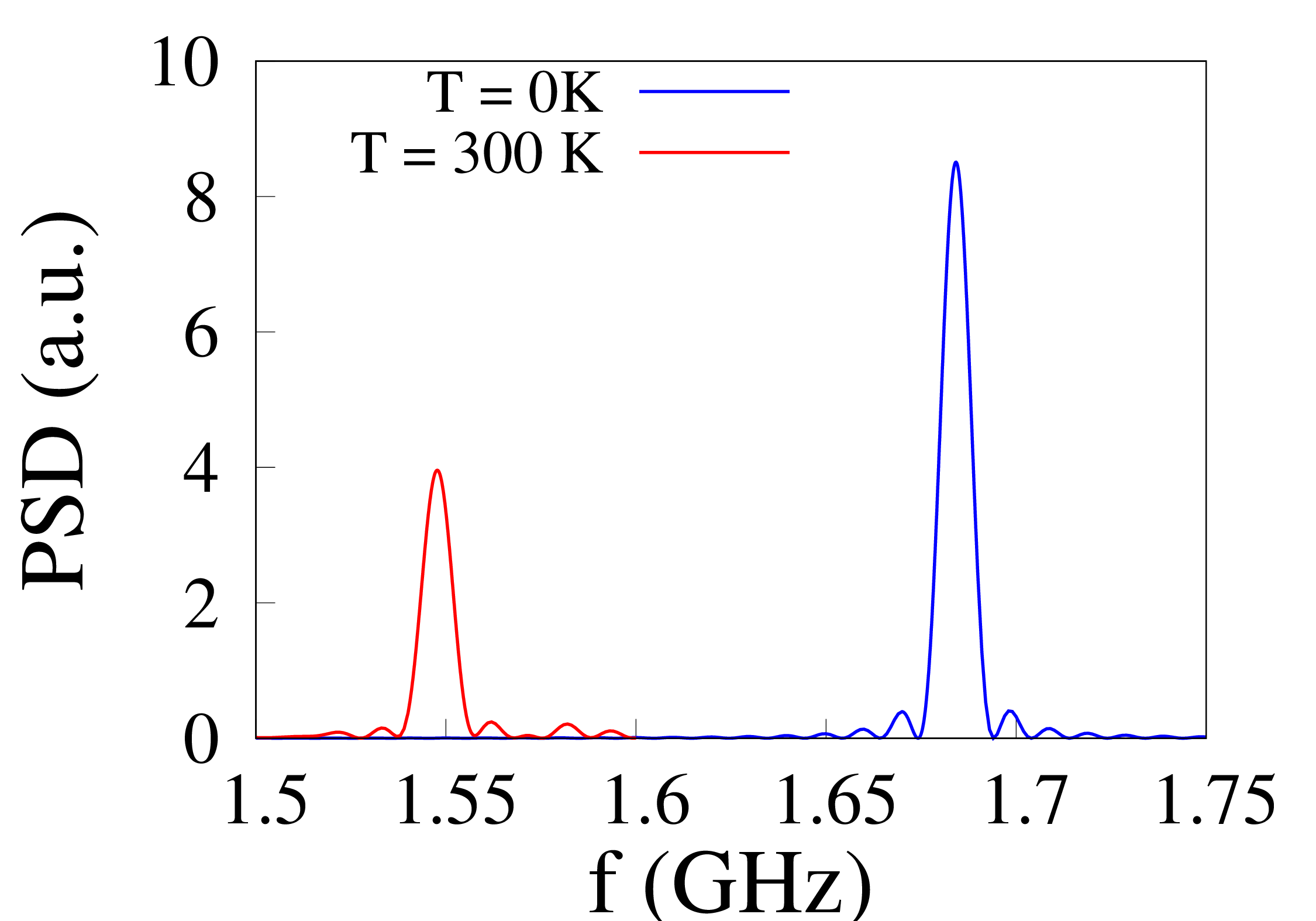}
		\caption{PSD for the temperatures $T$ = 0 K and 300 K corresponding to $I$ = 8 mA and $\beta$ = -0.6.}
		\label{Fig9}
	\end{figure}

To compare the power between the two different temperatures $T$ = 0 K and 300 K, we have plotted the PSD for $I$ = 8 mA and $\beta$ = -0.6 in Fig.\ref{Fig9}. The figure implies that the frequency is reduced from 1.68 GHz to 1.54 GHz when the temperature is increased from 0 K to 300 K.  The Q-factor is calculated as 191.39 and 172.35, respectively, from the linewidths 8.80 MHz and 8.98 MHz, respectively.

\section{Conclusion}
In conclusion, the emergence of self-oscillations in magnetic double tunnel junctions due to self-induced torque has been investigated numerically by solving the LLGS equation.  It was shown that the SIT induces the magnetization to precess continuously around the perpendicular direction of the film plane while only magnetization switching is exhibited without the SIT.  In the absence of FLT the frequency of oscillations gradually decreases with the current while the power gets enhanced. The presence of FLT considerably modifies the frequency by increasing its tunability range. When the strength of FLT is negative the frequency as well as the range of oscillations increases with current and when it is positive the frequency decreases and the range increases.   The frequency exhibits almost linear variation with the current as well as the strength of the FLT $\beta$.  The rate at which the frequency enhances increases with the increase of $\beta$ with negative sign.  When the value of $\beta$ is positive there exist critical values of current and $\beta$ above which the system exhibits no oscillations but only switching.  The power spectral density confirms that the power and Q-factor increases with oscillation frequency while increasing the current with the negative $\beta$.

\section*{Acknowledgement}
R.A. would like to thank SRM TRP Engineering College, India, for their financial support, vide No. SRM/TRP/RI/005. The works of R.G. and V.K.C. are supported by the DST and ANRF under Grant No. CRG/2023/003505.
and they wish to acknowledge DST, New Delhi, for computational facilities under the DST-FIST programme (No. SR/FST/PS-1/2020/135) to the Department of Physics. M.L. wishes to acknowledge the ANRF award of an ANRF-SERB National Science Chair under Grant No. NSC/2020/00029.

\section*{Data availability statement}
The data that support the findings of this study are available
upon reasonable request from the authors.

\section*{Appendix A}
The dynamics of the magnetization in the free layer is governed by the associated LLGS equation along with the SIT and it is given by
	\begin{equation}
 \frac{d\bf m}{dt} = -\gamma ~ {\bf m}\times{\bf H}_{eff} + \alpha ~{\bf m}\times  \frac{d\bf m}{dt}+\gamma H_s~ {\bf m}\times({\bf m \times \bf p})+\gamma \beta H_s {\bf m}\times {\bf p}
	+\gamma ~G~ R ~{\bf m\times (m\times p)}. \tag{A1}\label{eq:sup1} 
\end{equation}
	Here,  
\begin{eqnarray}
	G &=& \mu H_s ,\nonumber\\
	{\bf p} &=& {\bf e}_z,\nonumber\\
	R &=& {\bf p}\cdot ({\bf m \times {\bf \dot{m}}}),\nonumber\\
	{\bf H}_{eff} &=& H_k m_z {\bf e}_z- 4\pi M_s m_z  {\bf e}_z, \nonumber\\
	\alpha &=& \alpha_0+\frac{\hbar^2\gamma}{4e^2 M_s W} \left[\tilde{g}_l+\tilde{g}_r\right],\nonumber\\
	H_s &=& \frac{\hbar  \eta I}{2 e M_s W(1+\lambda {\bf m\cdot p})},\nonumber\\
	\mu &=& -\frac{ \hbar g_l^s}{4eg_r V}.\nonumber
\end{eqnarray}
The quantities $\tilde{g}_l$, $\tilde{g}_r$, $g_l^s$ and $g_r$ are defined as
\begin{eqnarray}
	\tilde{g}_l &=& \frac{1}{2}[g_l^P+g_l^{AP}-(g_l^P-g_l^{AP}){\bf m\cdot p}],\nonumber\\
	\tilde{g}_r &=& g_r = g_l^P,\nonumber\\
	g_l^s &=& P^{-1}(g_l^P-g_l^{AP}).\nonumber
\end{eqnarray}
$V$ is the voltage across the system given by
\begin{equation}
	V = I \frac{R_l (R_0-\triangle R~ m_z)}{R_l+R_0-\triangle R~ m_z}, \nonumber
\end{equation}
where $\triangle R$ = 0.5 $R_P$ $G_{MR}$ and $R_0$ = $R_P + \triangle R$.

Eq.\eqref{eq:sup1} is numerically solved for the magnetization of the free layer for the parameters listed in Table.\ref{table1}.

Hence, $\mu$ is given by
\begin{align}
	\mu = -\frac{0.07}{0.12} \frac{\hbar}{4e P I}\left[\frac{1}{R_0-\triangle R~ m_z}+\frac{1}{R_l}\right].\nonumber
\end{align}
For the numerical simulation, Eq.\eqref{eq:sup1} is derived for its components as follows: \\
After making dot product with Eq.\eqref{eq:sup1} by ${\bf p}$, we get

\begin{align}
{\bf p \cdot {\bf {\dot m}}} = -\gamma {\bf p}\cdot ({\bf m}\times {\bf H}_{eff})+\alpha {\bf p}\cdot({\bf m}\times {\bf {\dot m}}) + \gamma H_s {\bf p}\cdot [{\bf m}\times{\bf m}\times {\bf p}]\nonumber\\
+\gamma\beta H_s {\bf p}\cdot ({\bf m}\times{\bf p})+\gamma \mu H_s R {\bf p}\cdot [{\bf m}\times ({\bf m}\times{\bf p})] \tag{A2}\label{eq:sup2}
\end{align}
After substituting ${\bf p}\cdot [{\bf m}\times ({\bf m}\times{\bf p})]= -|{\bf m}\times{\bf p}|^2$ and ${\bf p}\cdot ({\bf m}\times{\bf p})$ = 0, from \eqref{eq:sup2} we can obtain $R$ as
\begin{equation}
	R =  A {\bf p}\cdot {\bf \dot m}+B, \tag{A3}\label{eq:sup3}
\end{equation}
where
\begin{equation}
	A = (\alpha-\gamma H_s \mu |{\bf m}\times {\bf p}|^2)^{-1}, \nonumber
\end{equation}
\begin{equation}
	B = \gamma A [{\bf p}\cdot ({\bf m}\times {\bf H}_{eff})+H_s |{\bf m}\times {\bf p}|^2]. \nonumber
	\end{equation}

	After taking cross product with Eq.\eqref{eq:sup1} by ${\bf m}$ we can obtain
\begin{align}
	{\bf m}\times{\bf \dot m} = &-\gamma~ {\bf m}\times({\bf m}\times {\bf H}_{eff})+\alpha~ {\bf m}\times({\bf m}\times {\bf \dot m})+\gamma H_s~ {\bf m}\times [{\bf m}\times ({\bf m}\times {\bf p})]\nonumber\\ &+ \gamma H_s \beta~ {\bf m}\times({\bf m}\times {\bf p}) + \gamma \mu H_s R~ {\bf m}\times [{\bf m}\times ({\bf m}\times {\bf p})]. \tag{A4}\label{eq:sup4}
\end{align}
After substituting ${\bf m}\times ({\bf m}\times {\bf \dot m})= -{\bf \dot m}$ and ${\bf m}\times [{\bf m}\times ({\bf m}\times {\bf p})] = - {\bf m}\times {\bf p}$ in \eqref{eq:sup4} we can get
\begin{align}
	{\bf m}\times {\bf \dot m} = -\gamma {\bf m}\times ({\bf m}\times {\bf H}_{eff})-\alpha {\bf \dot m}+\gamma \beta H_s {\bf m}\times ({\bf m}\times {\bf p})-\gamma H_s {\bf m}\times {\bf p}-\gamma H_s \mu R {\bf m}\times {\bf p}.\tag{A5}\label{eq:sup5}
\end{align}

	After substituting Eq.\eqref{eq:sup5} and Eq.\eqref{eq:sup3} in Eq.\eqref{eq:sup1} we can obtain
	
\begin{align}
{\bf \dot m} = &-\gamma {\bf m}\times {\bf H}_{eff} - \alpha \gamma {\bf m}\times ({\bf m}\times {\bf H}_{eff})-\alpha^2 {\bf \dot m}+\alpha \gamma \beta H_s {\bf m}\times ({\bf m}\times {\bf p})\nonumber\\
&-\alpha \gamma H_s {\bf m}\times {\bf p}-\alpha \gamma H_s \mu R {\bf m}\times {\bf p}+\gamma H_s {\bf m}\times ({\bf m}\times {\bf p})\nonumber\\
&+\gamma \beta H_s {\bf m}\times {\bf p} + \gamma \mu H_s R {\bf m}\times ({\bf m}\times {\bf p})\tag{A6}\label{eq:sup6}.
\end{align}
Eq.\eqref{eq:sup6} can be written as,
\begin{align}
	(1+\alpha^2) {\bf \dot m} =  &-\gamma {\bf m}\times {\bf H}_{eff} - \alpha \gamma {\bf m}\times({\bf m}\times {\bf H}_{eff}) +\gamma H_s (1+\alpha\beta){\bf m}\times({\bf m}\times {\bf p})\nonumber\\
	&+\gamma H_s(\beta-\alpha){\bf m}\times {\bf p} +\gamma H_s \mu   (A ~{\bf p}\cdot {\bf \dot m}+B)  [{\bf m}\times({\bf m}\times {\bf p})-\alpha{\bf m}\times {\bf p}]. \tag{A7}\label{eq:sup7}
\end{align}

After substituting ${\bf p}$ = ${\bf e}_z$ and ${\bf H}_{eff} = H_k m_z {\bf e}_z- 4\pi M_s m_z {\bf e}_z$ in Eq.\eqref{eq:sup7} the components form of the magnetization can be expressed as given below:
\begin{align}
&\frac{1+\alpha^2}{\gamma} { \frac{d m_x}{dt}} = \mu H_s A \frac{dm_z}{dt}(m_z m_x-\alpha m_y)-(H_k-4\pi M_s)m_z(m_y+\alpha m_z m_x) \nonumber\\&\hskip 2.5cm+H_s (1+\alpha \beta+\mu B)m_z m_x + H_s(\beta-\alpha-\alpha \mu B)m_y, \tag{A8}\label{eq:sup8}\\
&\frac{1+\alpha^2}{\gamma} \frac{d m_y}{dt} = \mu H_s A \frac{dm_z}{dt}(m_y m_z+\alpha m_x) + (H_k-4\pi M_s)m_z (m_x-\alpha m_y m_z)\nonumber\\&\hskip 2.5cm+H_s (1+\alpha \beta+\mu B)m_y m_z - H_s(\beta-\alpha-\alpha \mu B)m_x, \tag{A9}\label{eq:sup9}\\
&\left[\frac{1+\alpha^2}{\gamma}+\mu H_s A (1-m_z^2)\right] \frac{d m_z}{dt} = \alpha (H_k-4\pi M_s)m_z (1-m_z^2)\nonumber\\&\hskip 2.5cm-H_s (1+\alpha \beta+\mu B)(1-m_z^2), \tag{A10}\label{eq:sup10}
\end{align}
where $A = [\alpha -\gamma H_s \mu (1-m_z^2)]^{-1}$ and $B = A\gamma H_s(1-m_z^2)$.

\end{document}